\documentclass[twocolumn,showpacs,pra,aps]{revtex4-1}

\usepackage{dcolumn}
\usepackage{amssymb}

\draft

\begin{document}

\title{Dynamical Virial Relations \\ %
and Invalidity of the Boltzmann Kinetic Equation}

\author{Yu. E. Kuzovlev}
\email{kuzovlev@fti.dn.ua} \affiliation{Donetsk Institute
for Physics and Technology of NASU, 83114 Donetsk, Ukraine}


\begin{abstract}
A sequence of exact relations is found which connect %
one- and many-particle time-dependent distribution %
functions of low-density gas with their derivatives %
in respect to mean density. It is shown that, %
at least  in the context of spatially non-uniform %
gas evolutions, these relations forbid the ``molecular %
chaos propagation'' and imply inapplicability of the %
Boltzmann kinetic equation even under the %
Boltzmann-Grad limit and regardless of degree of the non-uniformity.
\end{abstract}


\maketitle

{\bf 1.\,Introduction}.\, %

\,\,\,

Subject of our interest here will be physical status of  %
the celebrated L.\,Boltzmann's kinetic equation (BE) %
for classical gas %
\cite{bol,bog,re,lp,bal,ub}. %
It can be written as
\begin{eqnarray}
\begin{array}{c}
\dot{F}(t,r,v)=-v\nabla F(t,r,v)\,+ \\ %
+\, \, n\,{\bf C}_2\,F(t,r,v)*F(t,r,w)\,\,, \label{be}
\end{array}
\end{eqnarray}
where\, $\,F(t,r,v)\,$\, is one-particle distribution function %
normalized to volume,\, \,$\,n\,$\, is mean gas density,\, and %
$\,{\bf C}_2\,$ is the Boltzmann's pair collision operator %
(``collision integral'') %
which acts onto $\,F\,$'s velocity argument as
\begin{equation}
\begin{array}{c}
{\bf C}_2\,F(v)*F(w)\,= \int d^3w \,|v-w| %
\int d^2b\, \times \\ \times\, %
 [F(v^{\,in}(b,v,w)) F(w^{\,in}(b,v,w))-F(v)F(w)]\, %
\label{bo}
\end{array}
\end{equation}
with\, $\,b\,$\, being the impact parameter vector %
(perpendicular to\, $\,v-w\,$) and\, $\,v^{\,in}\,$ and %
$\,w^{\,in}\,$\, input velocities what lead %
to the given output ones. %

Anybody can agree that today's kinetic theory, %
as well as 100 years ago, is unthinkable %
without BE. Nevertheless, nobody have presented %
a rigorous and at the same time general enough %
derivation of BE from such exact equations of %
statistical mechanics as the BBGKY equations %
(BBGKYE) \cite{bog,re,lp,bal}. %

In general, BBGKYE can only produce something like %
\begin{equation}
\begin{array}{c}
\dot{F}=-v\nabla F + %
n\,{\bf C}_2\,F*F  + %
n^2\,{\bf C}_3\,F*F*F +\, \\ %
\,+\,\,n^3\,{\bf C}_4\,F*F*F*F\,+\dots\, %
\end{array}
 \,\, \label{gbe}
\end{equation}
(see below).  %
Therefore, BE follows from BBGKYE %
under formal ``low-density %
limit'' only, when\, %
$\,n\rightarrow 0\,$\, %
\cite{bog,re,lp,bal}.
But this is non-physical %
limit, since it enforces the mean free path of %
gas atoms,\, $\,\lambda\sim 1/\pi a^2 n\,$ %
(with\, $\,a\,$\, denoting characteristic radius %
of atom-atom interaction), to tend to infinity.

Much more physically reasonable idealization is %
the ``Bpltzmann-Grad limit'' (BGL) %
when\, $\,n\rightarrow \infty\,$\, %
and simultaneously\, $\,a\rightarrow 0\,$\,, %
in such way that\, $\,\lambda\,$\, %
is kept constant. %
At that,\, $\,a^3n\rightarrow 0\,$\,, %
i.e. gas becomes ``infinitely rare''. %
This gave rise to hopes that the terms with\, %
$\,{\bf C}_3\,$, %
$\,{\bf C}_4\,$,\, etc., in Eq.\ref{gbe} %
must vanish under BGL too. %
Moreover, O.\,Lanford and others %
even suggested a proof %
of this hypothesis for hard-sphere %
gas \cite{lan,vblls}. %
But their interpretation of BBGKYE in this %
singular case was definitely wrong %
(in fact they from the very beginning substituted %
 interaction %
terms there by pair collision operators) %
\cite{hs1}.

In any case, %
contributions with\, $\,{\bf C}_3\,$, $\,{\bf C}_4\,$, %
etc., in Eq.\ref{gbe} appear due to %
violation of the Boltzmann's ``molecular %
chaos hypothesis'', that is due to %
statistical correlations between two atoms %
what are in mutually input ({\bf pre-collision}) %
states (approaching each other). %
Naturally, such correlations take place mainly %
at inter-atom separations comparable %
with \, $\,a\,$\,\,, %
but this does not mean that they vanish under BGL %
(merely they sit just where they are %
most ``harmful'').

What is physical meaning of the pre-collision %
correlations\,?

Notice that BE (\ref{be})-(\ref{bo}) %
fully corresponds to the ``probability-theoretical %
view'' at physical world:\, it presumes that %
gas evolution consists of elementary random events %
(collisions with various input velocities and impact parameters) %
which occur independently one on another and all possess %
strictly certain (though may be unknown numerically) %
\,{\it a priori}\, (conditional) probabilities. %

Such the view was originated by J.\,Bernoully 300 years ago \cite{jb}. %
But in 20-th century  N.\,Krylov \cite{kr} showed that %
assumption that any sort of events %
can be furnished with certain %
\,{\it a priori}\, probability, or relative frequency, has %
no support in rigorous statistical mechanics. %
In opposite, mechanics allows different relative frequencies %
on different phase trajectories of a many-particle system. %
Then, fluctuations of relative frequencies %
from one phase trajectory to another, - %
being  averaged over statistical ensemble of %
trajectories (experiments), - %
produce correlations between events and particles %
even in absence of %
cause-and-consequence connections between them.

Just such statistical correlations %
were found and investigated %
in my works \cite{i1} (see also \cite{i2}) and  %
\cite{p1,p0802,p0806,tmf,ig,p1105} %
on spatially inhomogeneous %
evolutions of fluids and random walks of their particles. %
And just such correlations %
determine higher-order terms in Eq.\ref{gbe}, %
impeding reduction of BBGKYE to BE %
even in spite of BGL. %

At that, role of the inhomogeneity  %
is to visualize and magnify %
effects of the uncertainty  of %
collision probabilities %
of fluid's particles. %
In particular experiments %
it manifests itself %
in the form of scaless (1/f\,-type) %
low-frequency fluctuations of %
both individual particles' mobilities %
and collective transport characteristics of a fluid. %

\,\,\,

Here, I suggest new proofs %
of invalidity of the BE (\ref{be}) in spatially %
inhomogeneous problems \cite{fn}. %
Starting from general properties and %
exact density expansions of %
\, $\,F(t,r,v)\,$\, and many-particle %
distribution functions (DF), %
we will derive a sequence of exact relations %
between them and their derivatives %
in respect to density \,$\,n\,$\,, %
and then, with the help of these relations, demonstrate  %
that the pre-collision inter-particle statistical correlations %
stay finite and significant even under BGL. %

\begin{widetext}

\,\,\,

{\bf 2.\,Many-particle statistical dynamics\,.}\, %

\,\,\,

Let $\,N\gg 1\,$ gas atoms %
are contained, %
by means of an auxiliary external potential, %
in volume\, $\,\Omega = N/n\,$\,. %
Let\,
$\,x_i=\{r_i,v_i\}\,$\, denote variables %
of\, $\,i\,$-th,\ atom,

$\,F_i(t)\equiv F(t,x_i)\equiv F_1(t,x_i)\,$\,,\, %
$\,F_{ij}(t)\equiv F_2(t,x_i,x_j)\,$\,,\, %
$\,F_{ijk}(t) \equiv F_3(t,x_i,x_j,x_k)\,$\,,\, %
... \, %
$\,F_{1\dots N}(t) \equiv %
F_N(t,x_1,x_2,\,\dots\, x_N)\,$\,
are one-, two-, %
three-,\, ... \, $\,N\,$-particle DFs,\,

$\,L_i\,$\,,\, %
$\,L_{ij}\,$\,,\, %
$\,L_{ijk}\,$\,,\, ... \,  %
$\,L_{1\dots N}\,$\, %
are one-, two-, %
three-,\, ... \, $\,N\,$-particle %
Liouville operators (including the %
auxiliary potential),\, and

$\,S_i(t) =  \exp{(L_i \,t)}\,$\,,\, %
$\,S_{ij}(t) = \exp{(L_{ij} \,t)}\,$\,,\, %
... \, %
$\,S_{1\dots N}(t) = \exp{(L_{1\dots N} \,t)}\,$\,
\,are corresponding evolution operators. %

All these objects are completely symmetric functions %
of their arguments (indices).

If all DF are thought normalized to volume %
and all obey the requirement %
of mutual consistency, %
then
\begin{eqnarray}
\frac 1{\Omega^{s}} \int_{1}\dots %
\int_s F_{1\,\dots\,s}(t)\,\,=\,1 \,\,\,,
\,\,\,\,\,\, %
F_{1\,\dots\,s}(t)\, \,=\, %
\frac 1{\Omega} \int_{s+1} %
F_{1\,\dots\,s\, s+1}(t)\,=\, %
\frac 1{\Omega^{N-s}} \int_{s+1}\dots %
\int_N F_{1\,\dots\,N}(t)\,\,\,,
\label{nc}
\end{eqnarray}
where\, $\,\int_k \dots \,\equiv \, \int_{\Omega} %
d^3r_k \int d^3v_k \dots \,$\,.\, %
Evolution of all marginal DFs is determined by that %
of the whole system's DF according to %
\begin{equation}
\begin{array}{c}
F_{1\,\dots\,N}(t)\, \,=\,  %
S_{1\,\dots\,N}(t)\, %
F_{1\,\dots\,N}(0)\,\,\, \label{ev}
\end{array}
\end{equation}
Since it conserves full phase volume, %
the equalities (\ref{nc}) will be satisfied at all %
times if it is so at one, ``initial'', %
time moment, e.g. at $\,t=0\,$. %
But some specific form of DFs can realize at one %
moment only. For example, we may choose
\begin{equation}
\begin{array}{c}
F_{1\,\dots\,s}(0)\, \,=\,  %
\prod_{j=1}^s\, F_j(0)\,\,\,, \label{ch}
\end{array}
\end{equation}
thus assuming ``molecular chaos'' at initial time moment.

Further, it is convenient to introduce %
operation \,$\,\circ\,$\, of %
``coherent product''  %
of the evolution operators. %
By definition, for any two  %
non-intersecting sets of indices, %
\begin{equation}
\begin{array}{c}
S_{i\dots j}\,\circ\,\, S_{k\dots \,l} \,=\, %
S_{i\dots j\,k\dots \,l}\,\,\, \label{cp}
\end{array}
\end{equation}
In essence, the left side here is mere equivalent %
notation for the right-hand side. %
For intersecting sets, of course, one must %
take their union. %
Thus, for instance,\, %
$\,S_{i\dots j}= S_i\circ \dots %
\circ S_{j} \,$\,,\, %
and we can write identities
\begin{eqnarray}
S_{1\dots N}\,  \,=\, S_{1\dots s} %
\prod_{j\,=\,s+1}^N \circ S_j\,=\, %
S_{1\dots s} %
\prod_{j\,=\,s+1}^N %
\circ [\,1\,+\,(S_j -1)\,]\,=\, %
\label{sid}  %
\\ =\, %
S_{1\dots s}\,+\, %
\sum_{j\,=\,s+1}^N \, S_{1\dots s} \circ (S_j-1) %
\,+\, \sum_{s+1\,\leq j<k\,\leq N}\, %
S_{1\dots s} \circ (S_j-1) \circ (S_k-1)\,+\,  %
\nonumber \\  +\, %
\sum_{s+1\,\leq j<k<l\,\leq N}\, %
S_{1\dots s} \circ (S_j-1) \circ (S_k-1) %
\circ (S_l-1)\,+\,\dots \,+\,  %
S_{1\dots s} %
\prod_{i\,=\,s+1}^N %
\circ (S_i -1)\, =\,\, %
\nonumber \\ =\, %
S_{1\dots s}\,+\, %
\sum_{j\,=\,s+1}^N \,[\,S_{1\dots s\,j} %
- S_{1\dots s}\,]\,+\, %
\sum_{s+1\,\leq j<k\,\leq N}\,%
[\,S_{1\dots s\,j\,k} %
- S_{1\dots s\,j} - S_{1\dots s\,k} %
+ S_{1\dots s} \,]\,+\, \dots \,\,\,, \nonumber
\end{eqnarray}
where\, $\,0\leq s<N\,$ (at\, $\,s=0\,$\,, %
of course, both\, $\,S_{1\dots s}\,$\, %
and\, $\,F_{1\dots s}\,$\, %
should be replaced by \,1\,). %

Combining these identities with Eqs.\ref{nc} %
and \ref{ev}, after simple algebra one obtains %
\begin{eqnarray}
F_{1\dots s}(t)\,=\, %
S_{1\dots s}(t)\, F_{1\dots s}(0)\,+ %
\sum_{k\,=\,1}^{N-s}\, %
\frac {(N-s)!}{k!\,(N-s-k)!\,\Omega^k} %
\int_{s+1}\dots \int_{s+k} %
\{\, S_{1\dots s}(t) \prod_{j=1}^k\, %
\circ [\,S_{s+j}(t)-1\,]\,\} \,\, %
F_{1\,\dots\,s+k}(0)\,\,\,\,\,\,\, \label{fid0}
\end{eqnarray}

Advantage of this representation %
of the marginal DFs' evolution,  %
in comparison with that given by %
(\ref{nc}) and (\ref{ev}), %
is that it can be directly extended %
to the {\bf thermodynamic limit}, %
when\, $\,N\rightarrow\infty\,$ and %
$\,\Omega\rightarrow\infty\,$\, %
(the auxiliary potential removes)
at fixed\, %
$\,N/\Omega\rightarrow n=\,$const\,. %

\,\,\,

\,\,\,

{\bf 3.\,Density expansion}.\, %

\,\,\,

The Eq.\ref{fid0} by itself does prompt conditions %
of this extension. %
Namely, a limit form %
of the initial DFs \,$\,F_{1\dots\,s}(0)\,$\,  %
at\, $\,\Omega\rightarrow\infty\,$\,
should be such that all the integrals in Eq.\ref{fid0} %
turn to well converging ``proper integrals''.  %
That is such that, %
at any\, $\,s\,$ and\, $\,k\,$\,, %
\begin{equation}
\begin{array}{c}
[\,S_{s+1}(t)-1\,]\circ\, \dots \, %
\circ\, [\,S_{s+k}(t)-1\,]\,\, %
F_{1\,\dots\,s\,s+1\,\dots \,s+k}\, %
\rightarrow\, 0\,\, \, \,\,\,\,\, %
\texttt{when} \,\,\,\,\,\,\, %
r_{s+1}\,\dots\,r_{s+k}\,\, \rightarrow\,\infty %
\,\,\,,  \label{cond}
\end{array}
\end{equation}
in  fast enough (integrable) way. %
This means that ``at infinity'' %
(asymptotically) all the initial DFs are invariant %
in respect to both translations %
and collisions of atoms. %
It will be the case %
if our gas is thermodynamically equilibrium %
(hence, homogeneous and rest) at infinity, i.e.
\begin{equation}
\begin{array}{c}
F_{1\,\dots\,s-1\,s}\, %
\rightarrow\, F_{1\,\dots\,s-1}\, F^{(eq)}(v_s) \, %
\,\,\, \, \,\,\,\,\, \texttt{at} \,\,\,\,\,\,\, %
r_s\,\rightarrow\,\infty \,\,\,,  \label{n}
\end{array}
\end{equation}
where\, $\, F^{(eq)}(v)= (2\pi T/m)^{-3/2} %
\,\exp{(- mv^2/2T)}\,$\, %
is the Maxwell velocity distribution %
with some temperature\, $\,T\,$\, %
(\, $\,m\,$\, is atomic mass). %
This agrees with requirements (\ref{nc}) %
in their limit form, that is, - %
in the case (\ref{ch}), - with  %
\begin{eqnarray}
\lim_{\Omega\rightarrow\infty}\, %
\frac 1{\Omega} \int_{\Omega} d^3r_1 %
\int d^3v_1 F_1\,\,=\,1 \,\,\, \label{nc1}
\end{eqnarray}

We see that  %
the thermodynamic limit in Eq.\ref{fid0} %
presumes that our gas (or, generally, fluid) %
is essentially inhomogeneous %
(if not equilibrium at all). %

Then Eq.\ref{fid0} transforms to infinite series 
\begin{eqnarray}
F_{1\dots s}(t)\,=\, %
S_{1\dots s}(t)\, F_{1\dots s}(0)\,+\, %
\sum_{k\,=\,1}^{\infty}\, %
\frac {n^k}{k!} %
\int_{s+1}\dots \int_{s+k} %
\{\, S_{1\dots s}(t) \prod_{j=1}^k  %
\circ [\,S_{s+j}(t)-1\,]\,\} \,\, %
F_{1\,\dots\, s+k}(0)\,=\,\,
\label{fid}\\
=\,S_{1\dots s}(t)\, F_{1\dots s}(0)\,+\, %
n \int_{s+1} [\,S_{1\,\dots\,s\,s+1}(t)- %
S_{1\dots s}(t)\,]\, F_{1\,\dots\, s\,s+1}(0)\,+\, %
\nonumber\\ \,+\,\, %
\frac {n^2}{2!} \int_{s+1} \int_{s+2} %
[\,S_{1\,\dots\,s\,s+1\,s+2}(t)- %
S_{1\,\dots \,s\,s+1}(t) - %
S_{1\,\dots \,s\,s+2}(t) + %
S_{1\,\dots \,s}(t)\,]\,
F_{1\,\dots\, s\, s+1\,s+2}(0)\, +\,\,\dots %
\,\,\, \nonumber
 \end{eqnarray}
with initial DFs (and, hence, %
arbitrary-time DFs) satisfying %
conditions (\ref{cond}) and (\ref{n}), %
and\, $\,\int_j \dots \,= %
\int d^3r_j \int d^3v_j \dots \,$\,.

The Eq.\ref{fid} represents density expansion %
of total evolution operator of the %
infinitely-many-particle system. %
If its initial state is treated  %
irrespective to\, $\,n\,$\,, then  %
Eq.\ref{fid} becomes density expansion %
of its time-dependent future (or past) DFs. %
This formal  treatment is %
physically adequate at least %
when Eq.\ref{fid} applies (as below) to %
sufficiently rare gas, in particular, characterized by %
initial ``molecular chaos'' (\ref{ch}). %

\,\,\,

\,\,\,

{\bf 4.\,Dynamical virial relations\,.}\,\,

\,\,\,

Below, we will confine ourselves by dilute gas, %
starting from the ``molecular chaos'' %
and being interested in the issue %
of its ``propagation'' with time. %

At\, $\,s=1\,$\, and\, $\,s=2\,$\, %
Eq.\ref{fid}, as combined with (\ref{ch}), yields %
\begin{eqnarray}
F_1(t)\,=\, S_1(t)\, F_1(0)\,+\, \label{f1} %
n \int_{2} [\,S_{12}(t)- S_1(t)\,]\, F_1(0)\, F_2(0) %
\,\,+\, \\ \,+\,\, %
\frac {n^2}{2!} \int_2 \int_3 [\,S_{123}(t)- %
S_{12}(t) - S_{13}(t) + S_1(t)\,]\, %
F_1(0)\, F_2(0)\,F_3(0)\,  \, +\,\dots \,\,\,, %
\nonumber \\ %
F_{12}(t)\,=\, S_{12}(t)\, %
F_1(0)\, F_2(0)\,+\, \label{f2} %
n \int_{3} [\,S_{123}(t)- S_{12}(t)\,]\, %
F_1(0)\, F_2(0)\, F_3(0)\, %
\,+\, \\ \,+\,\, %
\frac {n^2}{2!} \int_3 \int_4 [\,S_{1234}(t)- %
S_{123}(t) - S_{124}(t) + S_{12}(t)\,]\, %
F_1(0)\, F_2(0)\,F_3(0)\, F_4(0)\,  %
\, +\,\dots \,\,\, \nonumber %
\end{eqnarray}

There is definite relation %
between these expressions. %
To perceive it, let us %

(i) introduce functions %
\begin{eqnarray}
G_{1\,\dots\,s}(t)\,\equiv\, %
\prod_{j=1}^s \circ [S_{j}(t) -1] 
\, \,F_{1\,\dots\,s}(0)\,+ %
\sum_{k\,=\, 1}^{\infty}\, \frac {n^k}{k!} %
\int_{s+1} \!\dots \!\int_{s+k} %
\prod_{j=1}^{s+k} \circ [S_{j}(t) -1] %
\,\, F_{1\,\dots\,s+k}(0) \,\,\, %
\label{g}
 \end{eqnarray}

(ii) return to Eq.\ref{fid},\, %
expend there factors\, %
$\,S_{1\,\dots\,s}(t)\,$ into ``coherent product'' %
of\, $\,1+[S_j(t) -1]\,$\,,\, %

(iii) take into account identity %
\begin{eqnarray}
0\,=\, %
\sum_{k\,=\,1}^{\infty}\, \frac {n^k}{k!} %
\int_{1}\dots \int_{k}\, %
[\,S_{1}(t)-1\,]\circ \, \dots \,\circ %
[\,S_{k}(t)-1\,] %
\,\, F_{1\,\dots\, k}(0) \,\,\, %
\label{g0} %
\end{eqnarray}
what follows from Eqs.\ref{fid0} and %
\ref{fid} at\, $\,s=0\,$\, %
(and seems trivial in view of general %
properties of Liouville operators),\, %
and %

(iv) with the help of (\ref{g0}) %
rewrite\, $\,F_1(t)\,$\, %
and\, $\,F_{12}(t)\,$ in the form %
\begin{equation}
\begin{array}{c}
F_1(t)\, =\, F_1(0)\,+\,G_1(t)\,\,\,, \\ %
\label{f12} %
F_{12}(t)\, =\, F_{12}(0)\,+\, %
F_1(0)\,G_2(t)\, +\, F_2(0)\,G_1(t)\, +\, %
G_{12}(t) \,\,\, %
\end{array}
\end{equation}

Next, let us apply here the %
initial ``molecular chaos'' (\ref{ch}) %
and consider the initial %
one-particle DF (normalized to volume according %
to (\ref{nc1})) as independent %
on mean gas density\, $\,n\,$\,. %
Then, - treating (\ref{g}) as functions %
of\, $\,n\,$\, too, - we obviously can write %
exact relations
\begin{eqnarray}
\frac {\partial}{\partial n}\,\,\,  %
G_{1\,\dots\,s}(t)\,=\, %
\int_{s+1} G_{1\,\dots\,s\,s+1}(t)\, \,\, \label{dg}
\end{eqnarray}
Besides, %
differentiations of (\ref{g0}) %
in respect to\, $\,n\,$\, produce identities %
\begin{equation}
\begin{array}{c}
\int_1\dots \int_s \, G_{1\,\dots\,s}(t)\,=\,0\, %
\,\,, \label{gz} %
\,\,\,\,\,\,\,\, \,\,\, %
\int_1\dots \int_k  \,\{\prod_{j=1}^k \circ %
[\,S_j(t)-1\,]\,\} \,\prod_{j=1}^k\, F_j(0)\, %
\,=\,0\, \,\, %
\end{array}
\end{equation}

At\, $\,s=1\,$\, equality (\ref{dg}), - %
together with Eqs.\ref{f12} and \ref{gz}, - gives  %
\begin{eqnarray}
\frac {\partial}{\partial n}\,\,\, %
F_1(t)\, =\, \int_2 %
[\,F_{12}(t) - F_1(0)G_2(t) - %
G_1(t)F_2(0) - F_1(0)F_2(0)\,]\, %
=\, %
\nonumber \\ \,=\, %
\int_2 %
[\,F_{12}(t) + G_1(t)G_2(t) - %
- F_1(t)F_2(t)\,]\,\,=\, %
\int_2 \,[\, F_{12}(t) - F_1(t)F_2(t)\,]\, \,\,\,\, %
\label{df} \,\,\,\,\,\,\,
\end{eqnarray}

Formulae (\ref{dg}) and (\ref{df}) %
are direct analogues of the %
``virial relations'' found and considered in %
\cite{p0802,p0806,tmf,ig,p1105} %
as inportant statistical properties of
``molecular Brownian motion''. %
Here, I added the word ``dynamical'' %
in order to underline their principal difference from %
the relations known in equilibrium %
statistical mechanics. %
Another form of the ``dynamical virial relations'' %
(DVR) (\ref{dg}) is expounded in Appendix 1 below. %

\,\,\,

\,\,\,

{\bf 5.\,Symbolic kinetic equation\,.}\,\,

\,\,\,

Kinetic equation (KE) is an equation what %
closely expresses, - like Eqs.\ref{be} and \ref{gbe}, -
time derivative of the one-particle %
DF through it itself \cite{bog}. %
To construct a KE, first let us write
\begin{equation}
\begin{array}{c}
L_{1\,\dots\,s}\,=\,L_{1\,\dots\,s}^0\, +\, %
L_{1\,\dots\,s}^\prime\,\,\,, \\ %
\label{l} %
L_{1\,\dots\,s}^0\,=\, %
\sum_{1\leq i \leq s}\, L_i^0\,\,\,, \,\,\,\,\,\,\, %
L_{1\,\dots\,s}^\prime\,=\, %
\sum_{1\leq i<j \leq s}\, %
L_{ij}^\prime\, \,\,,
\end{array}
\end{equation}
where operators\, $\,L^0\,$\, and\, %
and $\,L^\prime\,$\, represent free motion %
of atoms and their (pair) interactions, respectively: %
\[
L_i^0\,=\, - \,v_i\,\nabla_i\,\,\,, \,\,\,\,\,\,\,\, %
L_{ij}^\prime\,=\, %
m^{-1}\,\Phi^\prime (r_i-r_j)\, %
(\partial/\partial v_i\, -\, %
\partial/\partial v_j)\,\,\,,
\]
with\, $\,\Phi(\rho)\,$\, being (short-range repulsive) %
interaction potential and\, %
$\,\Phi^\prime (\rho) = \partial %
\Phi(\rho)/\partial \rho\,$\,. %
At once, introduce free evolution operators %
and ``scattering operators'':
\begin{equation}
\begin{array}{c}
S_{1\,\dots\,s}^0(t)\,\equiv\, %
\exp{(L_{1\,\dots\,s}^0\, t)}\, %
\,\,, \,\,\,\,\,\,\, %
Z_{1\,\dots\,s}(t) \,\equiv\, %
S_{1\,\dots\,s}(t)\,S_{1\,\dots\,s}^0(-t)\, \,\, \label{so}
\end{array}
\end{equation}
(taking in mind that for any of our evolution %
operators\, $\,S^{-1}(t) =S(-t)\,$). %

Then, perform time differentiation of Eq.\ref{fid}. %
Using  the boundary conditions at infinity, %
(\ref{cond}) or (\ref{n}), and the %
symmetry of DFs, is not too hard %
to  make sure that, naturally, the result %
is the BBGKYE:
\begin{eqnarray}
\dot{F}_{1\,\dots\,s}(t)\,=\, L_{1\,\dots\,s}\, %
F_{1\,\dots\,s}(t)\,+ %
\sum_{j=1}^s\, n\, \int_{s+1} %
L_{j\,s+1}^\prime \, F_{1\,\dots\,s+1}(t)\, %
\,\, \label{bbgky}
\end{eqnarray}
Hence, we have only to express\, $\,F_{12}\,$\, %
in terms of\, $\,F_{1}\,$.

With this purpose,  %
let us treat  Eq.\ref{f1}) %
as series expansion of\, $\,nF_1(t)\,$\, %
over\, $\,nF_1(0)\,$\, and try to invert it: %
\begin{eqnarray}
F_1(0)\,=\, S_1(-t)\, F_1(t)\, -\, %
n\, S_1(-t) \int_{2} [\,S_{12}(t)- S_1(t)\,]\, %
S_1(-t)\,S_2(-t)\,F_1(t)\, F_2(t)\, %
\,+\, \dots %
\,\,\, \label{f1i}
\end{eqnarray}
Inserting this into Eq.\ref{f2}, %
after a tedious non-commutative algebra, %
we obtain an infinite series whose %
first two terms look as follow, %
\begin{eqnarray}
F_{12}(t)\,= \,\sum_{k\,=\,0}^\infty\, n^k\, %
F_{12}^{(k)}(t)\,=\, \,\, \label{f2e}\\ \,=\, %
Z_{12}(t)\,F_1(t)\,F_2(t)\, +\, %
n \int_3 [\, Z_{123}(t)\,-\, Z_{12}(t)\,Z_{13}(t)\,-\, %
Z_{12}(t)\,Z_{23}(t)\, +\,Z_{12}(t)\,]\, %
F_1(t)\,F_2(t)\, F_3(t)\, \, %
+\, \dots\, \,\, \,\,\,\, \nonumber
\end{eqnarray}
Here,\, a term with $\,s\,$-th degree of one-particle %
DF is of\,  $\,(s-2)\,$-th order in respect to density %
and contains\, $\,s-2\,$\, integrations. %
Insertion of this expansion into Eq.\ref{bbgky} %
at\, $\,s=1\,$\, yields the Eq.\ref{gbe} with %
\begin{eqnarray}
{\bf C}_2\, F*F \,=\, \int_2 \, L_{12}^\prime\, %
Z_{12}(t)\,F*F\,\,\,, \label{c2}\\  %
{\bf C}_3\, F*F*F \,=\, \int_2\int_3 \, L_{12}^\prime\, %
[\, Z_{123}(t)\,-\, Z_{12}(t)\,Z_{13}(t)\,-\, %
Z_{12}(t)\,Z_{23}(t)\, +\,Z_{12}(t)\,]\,\, %
F*F*F\, \,\, \label{c3}
\end{eqnarray}
Spending a lot of time, one could continue series %
in (\ref{f1i}) and (\ref{f2e}) and find %
also\, $\,F_{12}^{(2)}(t)\,$\, %
and\, $\,{\bf C}_4\,$.

\,\,\,

Notice that two written out right-hand terms %
of Eq.\ref{f2e} confirm %
the results of semi-heuristic considerations %
\cite{bog,lp,msg}, %
although with essential differences. %
Namely, in \cite{lp} the second term is thought %
as a small correction to the first one, %
in the framework %
of the low-density limit (see Sec.1 above), %
and time arguments of the ``scattering operators'' %
there are not unambiguously defined. %
In contrast to it, %
time argument of our operators\, $\,Z_{ij}(t)\,$, %
$\,Z_{123}(t)\,$\,,etc., is definitely total %
evolution time, so that they represent %
the whole pre-history of collisions.

This difference is especially important from viewpoint %
of higher-order terms of Eq.\ref{f2e} %
as considered in the framework of such more %
adequate approximation as the BGL. %
Actually, one can see that lower-order terms %
in Eq.\ref{f2e} and hence in Eq.\ref{gbe} %
arise fully independently on higher-order terms %
in Eqs.\ref{f1}, \ref{f2} and \ref{f1i}. %
Therefore,  if higher-order terms %
of Eq.\ref{f2e} and \ref{gbe} were insignificant %
then this would mean that arbitrary long evolution of %
one-particle DF is determined by only a few collisions. %
Absurdity of this enforced deduction clearly %
shows that in fact any of (infinitely many) %
terms of Eq.\ref{f2e} essentially %
contributes to Eq.\ref{gbe} even under BGL. %
Hence, Eq.\ref{gbe} hardly can be useful %
in practice and sooner has a symbolic %
meaning only, while BE (\ref{be}) %
is invalid at all. %

Next, this logical necessity %
will be sustained mathematically. %

\,\,\,

\,\,\,

{\bf 6.\,Violation of ``molecular chaos propagation''}.\,\,

\,\,\,

The first term of the series Eq.\ref{f2e} %
practically reproduces the %
Bogolyubov's formulation \cite{bog} %
of the ``molecular chaos'' (MC) hypothesis:\, %
$\,\,F_{12}(t)\Rightarrow F_{12}^{(0)}(t) %
=Z_{12}(t)\,F_1(t)F_2(t)\,$\,. %
Or, exploiting the pair correlation function,\, %
$\,C_{12}(t)= F_{12}(t)-F_1(t)F_2(t)\,$\, %
(see Appendix 1), %
$\,C_{12}(t)\,\Rightarrow\,[Z_{12}(t)-1] %
\,F_1(t)F_2(t)\,$\, %
What does it say?

By definition of the scattering operators, %
\begin{equation}
\begin{array}{c}
Z_{12}(t)\, F(r_1,v_1)\, F(r_2,v_2) \,= \, %
F(r_1^\prime +v_1^\prime t,v_1^\prime )\, %
F(r_2^\prime +v_2^\prime t,v_2^\prime )\,\,\,,
\label{z12}
\end{array}
\end{equation}
where %
the primed variables represent %
such past state, time\, $\,t\,$\, %
ago, which lead to the given current state. %
Therefore, %
the MC hypothesis, - i.e. assumption about %
insignificance of second and higher terms of Eq.\ref{f2e}, -  %
impliess that\, $\,C_{12}(t)\,$\, %
 differs from zero for two sorts of states. %
First, for currently interacting atoms, %
at\, $\,|r_2-r_1|\lesssim a\,$\,. %
Second, for atoms in %
mutually post-collision (out-) states, %
at\, %
\,$\,v_{12}\cdot r_{12} >0\,$\,, %
$\,a\lesssim |r_{12}|< |v_{12}|t\,$\, %
and\, $\,|b|\lesssim a\,$\,, %
where\, %
$\,r_{12}=r_2-r_1\,$\,,\, $\,v_{12}=v_2-v_1\,$\,, %
and \,$\,b = r_{12}- v_{12} %
(v_{12}\cdot r_{12})/|v_{12}|^2\,$\, %
is impact parameter vector %
(already mentioned in Sec.1). %

Thus, under the MC there are no correlations %
between atoms in mutually pre-collision states, - %
defined like the post-collision ones but with\, %
\,$\,v_{12}\cdot r_{12} <0\,$\,, - %
and no correlations between mere close %
though non-interacting atoms %
(for which\, $\,|r_{12}|\,$\, is greater %
than\, $\,a\,$\, but comparable with\, $\,a\,$), %
but the payment for such pleasure is presence of the %
unreservedly far propagating  %
post-collision correlations. %

If it was really so under BGL, %
then the exact relation (\ref{df}), - %
after its multiplying by\, $\,n\,$\,, - %
would reduce to %
\begin{eqnarray}
n\, \frac {\partial}{\partial n}\,\,\, %
F_1(t)\, =\, n \int_2 %
[\,Z_{12}(t) -1\,]\, F_{1}(t)\, F_{2}(t)\, %
\label{dfa} \,\,
\end{eqnarray}
The multiplication ensures finiteness of both sides here %
under transition to BGL, along with transition from %
\,$\,n\,$\, to physically more meaningful variable %
like\, $\,\kappa =\pi a^2n = 1/\lambda\,$\,, %
so that \,$\,n\, \partial/\partial n\, \Rightarrow %
\,\kappa\, \partial /\partial \kappa\,$\,. %

In combination with Eq.\ref{z12} %
this equality yields %
\begin{eqnarray}
n\, \frac {\partial}{\partial n}\,\,\, %
F_1(t,r_1,v_1)\, =\, n \int_0^t d\tau %
\int d^3v_2 \int d^2b \,\, %
|v_1-v_2|\,\, %
[\,F_1(t,r_1 + (v_1^\prime -v_1)\tau, %
v_1^\prime )\, %
F_1(t,r_1 + (v_2^\prime -v_1)\tau, %
v_2^\prime )\,\,-\, %
\nonumber\\ \, %
-\, F_1(t,r_1, v_1)\, %
F_1(t,r_1 + (v_2 -v_1)\tau, v_2)\,]\,=\, %
\nonumber\\ \,=\, %
 n \int_0^t d\tau \,\,\exp{(-\tau v_1\nabla_1)} \, %
{\bf C}_2\, F_1(t,r_1 + v_1\tau, v_1)\, %
F_1(t,r_1 + v_2\tau, v_2)  \, \,\,\, %
\label{dfc}
\end{eqnarray}
Here\, $\,^\prime\,$\, plays the same role %
as in the BE (\ref{be}). %
And, as in (\ref{be}), transition to BGL
allows to neglect $\,F_1\,$'s %
changes at time and space scales %
related to\, $\,a\,$\,. %

Simultaneously, %
according to the previous section,  %
the same MC assumption, - that\, %
$\,F_{12}(t)=Z_{12}(t)\,F_1(t)F_2(t)\,$\,, - %
produces the BE (\ref{be}) itself. %

Hence, if this is true assumption %
then Eqs.\ref{dfc} and \ref{be} should be %
compatible one with another. %
In fact, however,  %
they can not be satisfied simultaneously, %
except purely spatially homogeneous case %
when\, $\,\nabla_1 F_1(t)= 0\,$\, !\, \cite{fn1} %
To become convinced of this fact in detail, %
one may e.g. consider %
linearized evolution of small local or periodic %
perturbations of equilibrium state %
(\,$\,F^{eq}(v_1)\,$\, from above). %

\,\,\,

Consequently, %
contribution of the rejected terms of Eq.\ref{f2e}  %
into integral in DVR (\ref{df}) is on order of %
its value (see Eq.\ref{dfa}) under the MC assumption, %
and the latter, %
as applied to spatially inhomogeneous evolutions, %
is incompatible with the exact relation (\ref{df}) %
even in spite of the BGL. %

In other words, %
the ``molecular chaos propagation'' %
fails in spatially inhomogeneous case, %
so that the BE proves to be invalid %
even under BGL \,\cite{fn2,fn5}. %

\,\,\,

I would like to underline that %
all the aforesaid equally embrace %
the hard-sphere interaction. %

To finish the paper and exclude %
hopes to ``save'' BE, %
we will supplement just presented proof of %
its invalidity with short notes %
on the pre-collision correlations. %
Additional comments %
on related many-particle correlations %
and their influence onto\, $\,F_1(t)\,$'s\, %
evolution are placed to Appendix 2.

\,\,\,

\,\,\,

{\bf 7.\, Excess and pre-collision correlations}.\,\,

\,\,\,

Let us return to Eq.\ref{f2e} and %
consider functions
\begin{equation}
\begin{array}{c}
\Delta F_{12} (t)\,\equiv\, %
 Z_{12}(t)\, \Delta F_{12}^\prime (t)\,\equiv\,
F_{12}(t) \,-\, Z_{12}(t)\, %
F_1(t)\,F_2(t)\,\,= \, %
\sum_{k\,=\,1}^\infty\, n^k\, %
F_{12}^{(k)}(t)\,=\, %
\\ \,=\, %
n \int_3 \,[\, Z_{123}(t)\,- %
\, Z_{12}(t)\,Z_{13}(t)\,-\, %
Z_{12}(t)\,Z_{23}(t)\, +\,Z_{12}(t)\,]\, %
F_1(t)\,F_2(t)\, F_3(t)\, \, %
+\, \dots\, \,\, \label{ec}
\end{array}
\end{equation}
They characterize those part of %
statistical correlations between two atoms %
what is due to not their interaction between themselves %
but  common pre-history of their interactions %
(collisions) with the rest of gas. %
At that,\, $\,k\,$-th term of the sum %
represents connected chains (clusters) of %
at least \,$\,k+1\,$\, %
collisions (actual or virtual ones) %
conjointly involving \,$\,k+2\,$\, atoms. %
Statistical meaning of such ``excess'' %
(or ``historical'' \cite{tmf}) correlations %
was exhaustively explained in \cite{i1,i2} %
(see also Sec.1 above and notes in \cite{hs1,p1,ig,hs}).

It is clear that, first, if\, %
$\,\Delta F_{12} (t)\neq 0\,$\, %
somewhere in two-particle phase space, %
then\, $\,\Delta F_{12} (t)\neq 0\,$\, %
almost everywhere, since anyway there are %
many various clusters of collisions resulting in given %
current states of atoms 1 and 2. %

Second, undoubtedly\, %
$\,\Delta F_{12} (t)\neq 0\,$\, %
for post-collision states, %
since\, $\,\Delta F_{12} (t)\,$\, %
must compensate and stop nonphysical %
unrestricted propagation %
of the above mentioned post-collision correlations %
prescribed by the MC approximation\, %
$\,F_{12} (t)= F_{12}^{(0)}\,$\,. %
Hence, undoubtedly\, $\,\Delta F_{12} (t)\neq 0\,$\, %
almost everywhere, including the pre-collision states. %

This reasoning shows that appearance %
of pre-collision correlations along with %
other (``non-collision'') excess correlations  %
(at states for which\, %
\, $\,\Delta F_{12} (t)\approx \Delta %
F_{12}^\prime (t)\,$\, %
in Eq.\ref{ec}) is quite inevitable from %
physical point of view. %

Third, on the other hand, %
separation of atoms 1 and 2 %
must decrease a number of the clusters %
determining\, $\,C_{12}(t)\,$\, %
nearly proportionally to visual space angle %
\,$\,\o\,$\, %
of one of the atoms from viewpoint of another, %
\,$\,\o \sim \pi a^2/4\pi |r_{12}|^2\,$\,. %
The matter is that, since\, $\,F_{12}^{(k)}(t)\,$\, %
involves \,$\,k+1\,$\, (or more) collisions %
but contains only \,$\,k\,$\, integrations, %
one (or more)  of  \,$\,k\,$\, integration %
velocities is restricted in respect %
to its direction by a space angle %
\,$\,\sim \o\,$\,. %

By these reasons, we can propose %
the following rough fit\,: %
\begin{eqnarray}
C_{12} (t)\,=\, \frac {a^2} %
{a^2 + 4|r_2-r_1|^2}\,\, %
C_{12}^\prime (t)\,\,\,, \label{ce}
\end{eqnarray}
where function\, $\,C_{12}^\prime (t)\,$\, %
keeps non-zero under BGL %
and smmothly depends on\, $\,|r_{12}|\,$\,  %
at\, $\,|r_{12}|\gtrsim a\,$\,. %

Then DVR (\ref{df}) implies (under BGL) that
\begin{eqnarray}
\frac {a^2}4 \int d\o \int_0^\infty %
d|r_{12}| \int d^3v_2  \,\,\, %
C_{12}^\prime (t)\,\,=\, %
\pi \, a^2\,\, %
 \frac {\partial F_1(t)}{\partial \kappa}\, %
\,\,, \label{dff}
\end{eqnarray}
where\, $\,\o = r_{12}/|r_{12}|\,$\,. %

\end{widetext}

\,\,\,

We see that at any fixed distance %
\,$\,|r_{12}|\,$\, the excess correlations, - %
particularly, the pre-collision ones, - %
turn to zero under BGL. %
In this sense, the MC really takes place. %

But contribution of these correlations %
to the ``triple collision integral'' %
(\ref{c3}), to\, $\,{\bf C}_4\,$\,,\,...\,, %
and thus to the whole Eq.\ref{gbe} %
is determined by region \,$\,|r_{12}|\sim a\,$\, %
where all excess (pre-collision) %
correlations stay finite under BGL \cite{fn4}. %
Hence, in essence MC fails.

\,\,\,

And last remarks. %

\,\,\,

(i)\, Both the left-hand integral in Eq.\ref{dff} %
and right-hand derivative there define some %
(one and the same) characteristic length. %
Of course, it must be nothing but the\, %
$\,\lambda\,$\, (for it would be very strange\, %
if\,  $\,C_{12}(t)\,$\, was indifferent to %
\,$\,\lambda\,$\,). %
At the same time, obviously, %
any of constituent parts of\, %
$\,C_{12}(t)\,$\,, namely,\, %
$\,[Z_{12}(t)-1]F_1(t)F_2(t)\,$\, and\, %
$\,F_{12}^{(k)}(t)\,$\, ($\,k=1,2,\dots\,$), %
extends up to\, $\,|r_{12}|\sim v_0 t\,$\,, %
with\, $\,v_0\,$\, being characteristic velocity %
of gas atoms. %
Hence, indeed all these parts are required in order %
to introduce\, $\,\lambda\,$\, %
in place of\, $\,v_0t\,$\,.

\,\,\,

(ii)\, The reasonings and conclusions %
of this section (as well as %
that at end of Sec.5 and in Appendix 2) %
in no way rely on some %
measure of non-uniformity of gas state. %
Therefore, all they are equally valid, - %
and thus BE is equally invalid, - %
for arbitrary weakly non-uniform gas \cite{fn5,fn6}. %


\,\,\,

\,\,\,

{\bf 8.\,Conclusion}.\,\,

\,\,\,

Considering spatially inhomogeneous %
evolution of low-density gas, we introduced a %
non-standard representation of its time-dependent %
distribution functions (DF) in the form of their %
density expansion, \, and then %
exploited it to derive original %
exact ``dynamical virial relations'' (DVR) %
connecting DFs with their density derivatives. %
\,Then we applied DVR to analysis of behavior %
of two-particle correlation function %
under the Boltzmann-Grad limit (BGL), %
in order to examine validity of the %
``folklore'' opinion %
that evolution of one-particle DF under BGL %
exactly obeys the Boltzmann kinetic equation (BE) %
while many-particle DFs undergo the %
``molecular chaos propagation''. %

We showed that the corresponding %
approximate approaches to gas %
kinetics \cite{bog,re,lp,bal}, which seem well %
grounded in the ``low-density limit'', %
at the same time appear non-grounded under BGL, %
since contradict the DVR and thus fail, %
when applied to %
spatially non-uniform situations %
(independently on degree %
of the non-uniformity). %

This fact does not mean, of course, %
that idea of the Boltzmann collision operator %
(integral) is defective. This is excellent concept %
if one applies it to a separate collision. %
But it by itself is unable to comprise %
those inter-atom statistical correlations %
what arise from uniqueness (``non-ergodicity'' \cite{tmf}) %
of histories of collisions in particular experiments %
(see Sec.1 above and comments in %
\cite{i1,i2,p1,p0802,tmf,ig,p1105,eiphg,p0705}). %

One of ways to take into account all these %
correlations is the approach to correct solution of %
the BBGKY equations (the ``collisional approximation'') %
suggested in \cite{i1} (see also \cite{i2,p1}). %
In principle, this approach allows to consider %
a wide variety of phenomena and problems %
(``from molecular Brownian motion to shock waves'' %
\cite{p0705}). %

In \cite{i1,p1} this approach was used to investigate  %
statistical characteristics of molecular random walks %
in fluids (in particular, the related 1/f noise). %
The results then were confirmed from viewpoint of %
corresponding exact ``virial relations'' %
\cite{p0802,tmf,p1105}. %
Thus, a part of ``Augean stables'' was %
cleansed:\, some ancient prejudices, %
pointed out in \cite{kr},  like %
``molecular chaos'', were overcome  %
with substantial physical profit. %

Therefore, it seems reasonable to try to %
apply the mentioned approach %
(strengthened with the DVR) %
to gas (fluid) kinetics and hydrodynamics too. %
All the more so as their difference %
from Boltzmannian kinetics and classical %
hydrodynamics may be practically important %
even for weakly non-equilibrium %
(non-uniform) processes \cite{fn3}. %


\,\,\,

\,\,\,

{\bf Appendix 1.\, Correlation (cumulant) %
distribution functions and the DVR}.\,\,

\,\,\,

Let us introduce irreducible many-particle correlation, %
or cumulant, functions (CF) by
\begin{equation}
\begin{array}{c}
F_{12} \,= \,F_{1}F_{2} + C_{12}\,\,,  %
\,\,\,\,\,\, %
F_{123} \,= \,F_{1}F_{2}F_3 + %
\\ +\, C_{12}F_3 +  %
C_{23}F_1 + C_{13}F_2 + C_{123}\,\,\,, \label{cfs}
\end{array}
\end{equation}
and so on. %
Higher-order CFs can be defined with the help of %
generating functionals: %
\begin{eqnarray}
1\,+\sum_{s=1}^\infty F_s(t)\,\psi^s/s! \,=\, %
\nonumber \\  %
\,=\, \exp\, [\,F_1(0)\,\psi \,]\,\, %
\{\,1\,+\sum_{s=1}^\infty G_s(t)\, %
\psi^s/s!\,\}\, =\, \nonumber \\ %
\,=\, \exp\, [\,F_1(t)\,\psi \,+ %
\sum_{s=2}^\infty C_s(t)\, \psi^s /s!\,]\,\,,\,\, %
\label{gfs}
\end{eqnarray}
where, of course,\, %
\begin{eqnarray}
F_s(t)\,\psi^s \,=\, %
\int_1\dots \int_s F_{1\,\dots\, s}(t) %
\,\psi(x_1)\dots \psi(x_s)\,\,\, \nonumber
\end{eqnarray}
and so on. %
In these shortened notations, %
the DVR (\ref{dg}) altogether can be %
accumulated into single generating DVR %
\begin{eqnarray}
\frac {\partial }{\partial n}\,  %
\,\sum_{s=1}^\infty G_s(t)\, %
\frac {\psi^s}{s!}\,\,\, =\, %
\int dx \frac {\delta }{\delta \psi(x)}\,  %
\sum_{s=1}^\infty G_s(t)\, %
\frac {\psi^s}{s!}\,\,\, \,\,\,\label{dgs}
\end{eqnarray}
(to be supplemented with equality %
(\ref{g0}) and thus also (\ref{gz})). %
Obviously, this generating DVR is equivalent to %
\begin{eqnarray}
\frac {\partial C_{1\,\dots\,s}(t)}{\partial n}\,\,  %
=\, \int_{s+1} C_{1\,\dots\,s\,s+1}(t)\, \,\, %
\, \,\, (s\geq 2) \,\,, \,\,\,\,\,\, %
\label{dc}\\
\frac {\partial F_{1}(t)}{\partial n}\,\,  %
=\, \int_{2} C_{12}(t)\, \,\, \,\, %
\label{dc1}
\end{eqnarray}

Generalization of these DVR to dense gases (fluids), - %
when the ``initial molecular chaos'' (\ref{ch}) is too %
bad choice for initial DFs (since they must %
include effects of atom-atom interaction %
and thus depend on the mean %
density), - will be considered elsewhere.

\,\,\,

\,\,\,

{\bf Appendix 2.\,\, Many-particle correlations %
and falsity of the BE's ``derivations''}.\,\,

\,\,\,

According to the BBGKY equations (\ref{bbgky}) %
or the DVR (\ref{dg}) and/or  (\ref{dc1}), %
the pair pre-collision correlations %
arise in company with various many-particle ones. %
Their importance for correct approach %
to gas (fluid) kinetics was demonstrated %
already in \cite{i1} (see also \cite{i2,p1,tmf}). %
For one more demonstration, %
let us criticize the ``derivation'' %
of BE (\ref{be}) suggested in \cite{bal}. %

Assume, as there,  that the irreducible (``pure'') %
three-particle correlations (see Appendix 1), - %
described by  CF\, $\,C_{123}(t)\,$\,, -  %
can be neglected at\, $\,a^3n\ll 1\,$\, %
and hence under BGL. %
Then the second of Eqs.\ref{bbgky}, %
when written in terms of CFs, reduces to %
\begin{equation}
\begin{array}{c}
\dot{C}_{12}\,=\, L_{12}\, C_{12}\, +\, %
L_{12}^\prime\, F_1\,F_2\, =\, %
L_{12}^0\, C_{12}\, +\, %
L_{12}^\prime\, F_{12}\,\, \label{bale}
\end{array}
\end{equation}

Solving Eq.\ref{bale} (with zero initial condition) %
and inserting the result,\, %
$\,C_{12}\,$\,, %
into the first of Eqs.\ref{bbgky}, %
one can come (at\, $\,a^3n\ll 1\,$\,) 
to the BE \cite{bal}. %

If this is true derivation of BE, %
then it should be compatible with the DVR, %
at least, with Eq.\ref{df}  %
(i.e. Eq.\ref{dc1}). %
In fact, however, this is not the case. %
Indeed, integration of Eq.\ref{bale} over\, %
$\,x_2\,$\,, after multiplying it %
by\, $\,n\,$\,, yields
\begin{eqnarray}
\frac {\partial }{\partial t}\, %
\,n \int_2 C_{12}\,=\, %
L_1^0\, n\int_2 C_{12}\, +\, %
n\int_2 L_{12}^\prime\, F_{12}\, =\, %
\nonumber \\ \,=\, %
L_1^0\, n\int_2 C_{12}\, +\, %
\left[\,\frac {\partial }{\partial t}\,-\, %
L_1^0\,\right]\, F_1\,\,\,, %
\nonumber
\end{eqnarray}
where we used also the first of %
the BBGKY equations (\ref{bbgky}). %
Combining this equality with the DVR (\ref{df}), %
we come to equality %
\begin{equation}
\begin{array}{c}
[\,\partial /\partial t\,-\,L_1^0\,]\, %
[\,n\, \partial F_1/\partial n\,\, -\,F_1\,] %
\,=\, 0\,\,\,, \nonumber
\end{array}
\end{equation}
which certainly is wrong.

Hence, the above BF's ``derivation'', based %
on approximation (\ref{bale}), is erroneous. %
And, in order to get a correct description of %
\,$\,F_1\,$'s evolution, one should %
seriously think about role of %
many-particle correlations.

\,\,\,

\,\,\,



\begin{thebibliography}{27}
\bibitem{fn}
This communication is dedicated to 30-the anniversary %
of the first papers \cite{pjtf} (see also %
\cite{bk12,bk3}) where the idea of %
realistic ``non-ergodic'' random walks %
(of charge carriers), - which have no certain %
diffusivity and mobolity and do not obey  %
Boltzmannian kinetics %
(``quasi-Gaussian random walks \cite{i2}), - %
was suggested and mathematically expounded %
(about my recent considerations on this subject %
see \cite{p1008,eiphg}).

\,\,\,
\bibitem{fn2}
Notice that the %
modified formulation \cite{lp} of the ``molecular
chaos'' hypothesis, like %
\,$\,F_{12}(t)=Z_{12}(t_0)\,F_1(t)F_2(t)\,$\,, - %
with a fixed\, $\,t_0\,$\, (at\, $\,t>t_0\,$\,), - %
is worse than the above formulation, since %
appears incompatible with virial relation (\ref{df}) %
even in homogeneous case.

\,\,\,
\bibitem{fn5}
It is useful to add that %
the incompatibility of (both hypothetical) equations %
(\ref{dfa}) and (\ref{be}) %
is sufficent but not necessary condition %
for invalidity of one (or both) of them. %
Therefore, both they can be wrong %
for homogeneous gas too. %
This suspicion  is confirmed by argumentation %
presented in Sec.7 %
(another matter that under our conditions %
for existence of the thermodynamic limit, - %
see Sec.3, - homogeneous gas means %
merely equlibrium gas). %

\,\,\,
\bibitem{fn4}
Seemingly, possibility of such rather subtle circumstance %
was not foreseen by the authors of \cite{lan,vblls}, %
and this oversight caused subtle errors in their %
formal construction (see notes in \cite{hs1,i2,p0806,hs}). %

\,\,\,
\bibitem{fn6}
This circumstance highlights that the Boltzmann equation %
is from mathematical tale but not from real life:\, %
it describes ensemble of ``dice tosses'' %
(``coin tossings'') instead of ensemble of dynamical %
trajectories (which in fact appears to be irreducible %
to ``dice tosses'' \cite{p0802,tmf}).

\,\,\,
\bibitem{fn3}
This statement comes from %
\,(i) essential difference of our theory of %
self-diffusion of gas atms \cite{i1} from the %
classical one, and %
\,(ii) observation  %
that in case of (infinitezimally) weak non-equilibrium %
equations of our approach (``collisional approximation'') %
\cite{i1} are quite similar to the equations describing %
the self-diffusion (the only difference is that %
the place of the Boltzmann-Lorentz operators %
is occupied %
by the linearized Boltzmann operators). %
\end{thebibliography}
\end{document}